\documentclass[twocolumn,twoside,slac_two]{revtex4}
\usepackage{graphicx}
\usepackage{fancyhdr}
\newcommand{\be}{\begin{equation}}
\newcommand{\ee}{\end{equation}}
\newcommand{\bary}{\begin{eqnarray}}
\newcommand{\eary}{\end{eqnarray}}

\begin{document}

\title{SSC EMISSION  AS EXPLANATION OF THE  GAMMA RAY AFTERGLOW OBSERVED IN GRB 980923}

%

\author{N. Fraija, M. M. Gonz\'alez and  W. H. Lee}
\affiliation{Instituto de Astronom\'ia, Universidad Nacional Aut\'{o}noma de M\'{e}xico, Apdo. Postal 70-264, Cd. Universitaria, M\'{e}xico DF 04510}
%

\begin{abstract}
GRB 980923 was one of the brightest bursts observed by the Burst and Transient Source Experiment (BATSE). Previous studies have detected two distinct components in addition
to the main prompt episode, which is well described by a Band function. The first of these is a tail with a duration of $\simeq 400$s, while the second is a high-energy component lasting $\simeq 2$~s.
After summarizing the observations, we present a model for this event and conclude that the tail can be understood as the early gamma-ray afterglow from forward shock synchrotron
emission, while the high-energy component is described by the SSC emission from the reverse shock. The main assumption is that of a thick-shell case from highly magnetized
ejecta. The calculated fluxes, break energies, starting times and spectral index are all consistent with the observed values.
\end{abstract}

\maketitle

\thispagestyle{fancy}


\section{INTRODUCTION}

The most successful theory to explain GRBs and their afterglows is the
fireball model \cite{meszaros}. This model predicts an expanding
ultrarelativistic shell that moves into the external medium. When the
expanding shell collides with another shell (internal shocks) or
surrounding interstellar media (external shocks) gives rise to radiation
emission through the synchrotron and SSC processes. Besides when
the expanding relativistic shell encounters the external medium involves
two shocks: an outgoing shock (the forward shock) and other that
propagates into the ejecta (the reverse shock). When the forward shock
collides with ISM, electrons are accelerated up to relativistic energies.
The reverse shock heats up the shell's matter and accelerates electrons
when it crosses the shell. Now, although the contribution of the
synchrotron emission of reverse shock to the X-ray band could be small,
electrons in the reverse shock region can upscatter the synchrotron
photons (SSC process) up to higher energies. In this work we extend
the smooth tail work done by \cite{Giblin et  al.} and study synchrotron self-inverse
Compton radiation from a thick shell of reverse shock fireball to
explain the hard component of the GRB 980923.

\section{GRB 980923}

GRB 980923 was observed by BATSE on 1998 September 23 at
20:10:52 UT for 32.02 s. It was localized to 2340 with respect to the
pointing-axis direction of CGRO. In accordance to \cite{gon11}, the
event consists of three components \cite{sac11a}. The first
component is related to the typical prompt emission, the second one is
related to the smooth tail which lasts 400s and the last one is related to
the hard component which shows a high energy spectral component
extending up to Å 150 MeV and the power index  -1.44 $\pm$ 0.07. The
smooth tail was well described by \cite{Giblin et  al.} as the evolution of a
synchrotron cooling break in the slow-cooling regime at $t_0$=32 seconds
where the characteristic value of the power index was p= 2.4 $\pm$ 0.11.
However,  \cite{gon11} point out that the tail could begin before or at
least about 14s after the burst trigger and after of a short period of time
occurred the transition between fast to slow cooling.

\section{Dynamics of the forward and reverse shock} 

In a unified way between forward and  reverse shock, we compute  the energy range for synchrotron self-inverse Compton radiation from a thick shell of the reverse shock fireball to explain the hard component. The subscripts $\rm f$ and $\rm r$ refer throughout to the forward and reverse shock, respectively.

\subsection{Smooth tail from Synchrotron radiation forward shock}

For the forward shock, we assume that electrons are accelerated in the shock to a power law distribution of Lorentz factor $\gamma_e$ with a minimum Lorentz factor $\gamma_m$: $N(\gamma_e)d\gamma_e\propto \gamma_e^{-p}d\gamma_e$,
$\gamma_e\geq\gamma_m$ and  that constant fractions $\epsilon_{e,f}$ and $\epsilon_{B,f}$ of the
shock energy go into the electrons and the magnetic field, respectively. Then
\begin{eqnarray}
\gamma_{m,f}&=&\epsilon_{e,f}\biggl(\frac {p-2} {p-1}\biggr) \frac {m_p} {m_e}\gamma_f\cr
&=&524.6 \,\epsilon_{e,f}\,\gamma_{\rm f}
\end{eqnarray}
where we have used the value of $p=2.4\pm 0.11$ as was obtained by \cite{Giblin et  al.}.   Using the typical parameters given by \cite{bjo01},  we compute the typical and cooling frequencies of the forward shock synchrotron emission \cite{sar98} which are given by,
\bary\label{synforw}
\nu_{\rm m,f}&\sim& 1.9 \times 10^{19} \biggl(\frac{1+z}{2}\biggr)^{1/2}\,\biggl(\frac{\epsilon_{e,f}}{0.95}\biggr)^2\,\epsilon^{1/2}_{B,f,-5}\cr
&& E^{1/2}_{54}\,t^{-3/2}_{1}\,{\rm \ Hz}\cr
\nu_{\rm c,f}&\sim& 3.0 \times 10^{19}\biggl(\frac{1+z}{2}\biggr)^{-1/2}\,\biggl(\frac{1+x_f}{2.5}\biggr)^{-2}\cr
&&\epsilon^{-3/2}_{B,f,-5}\,n^{-1}_{f,0}\,E^{-1/2}_{54}\,t^{-1/2}_{1}\, {\rm \ Hz}\cr
&&\cr
F_{\rm max,f}&\sim& 2.2\times 10 \biggl(\frac{1+z}{2}\biggr)\,\epsilon^{1/2}_{B,f,-5}\,n^{1/2}_{f,0}\,D^{-2}_{28}\,E_{54}\,{\rm \,\mu Jy}\cr
t_{\rm tr,f}&\sim& 8.7 \biggl(\frac{1+z}{2}\biggr)\, \biggl(\frac{\epsilon_{e,f}}{0.95}\biggr)^2  \,\epsilon^{2}_{B,f,-5}\,n_{f,0}\,E_{54}\,{\rm s}
\eary

\noindent where the convention $Q_x=Q/10^x$ has been adopted in cgs units throughout this document unless otherwise specified. $t_{\rm tr,f}$ is the transition time, when the spectrum changes from fast cooling to slow cooling,  $D$ is the luminosity distance, $n_{f}$ is the ISM density, $t$ is the time of the evolution of the tail, $E$ is the energy, and the term $(1+x_f)$ was introduced because a once-scattered synchrotron photon generally has energy larger than the electron mass in the rest frame of the second-scattering electrons.  Multiple scattering of synchrotron photons can be ignored. $x_f$ is given by \cite{sar01} as:

\begin{equation}
x_f = \left\{ \begin{array} {ll} 
\frac{\eta \epsilon_{e,f}}{\epsilon_{B,f}}, & \mathrm{if \quad}
\frac{\eta \epsilon_{e,f}}{\epsilon_{B,f}} \ll 1, \\ 
\left(\frac{\eta \epsilon_{e,f}}{\epsilon_{B,f}}\right)^{1/2}, & \mathrm{if \quad}
\frac{\eta \epsilon_{e,f}}{\epsilon_{B,f}} \gg 1. 
\end{array} \right.
\end{equation}
where $\eta=(\gamma_{\rm c,f}/\gamma_{\rm m,f})^{2-p}$ for slow cooling and $\eta=1$ for fast cooling.\\
From eq. (\ref{synforw}), we observe directly that $\nu_{\rm m,f}  \leq \nu_{\rm c,f}$, the break energy $E_{\rm c,f}\sim 124.1\,$keV and  $t_{\rm tr,f}\sim 8.7$, implying also that the transition from fast to slow cooling  could take place on  very short  timescales, comparable to the duration of the burst.

\subsection{ X ray flare from thick shell  Reverse shock}
For the reverse shock, it is possible to obtain a simple analytic solution in two limiting cases, thin  and thick  shell,  \cite{sar95} by using a critical Lorentz factor $\Gamma_c$, 

\bary
\Gamma_c&=&\biggl(\frac{3E}{4\pi n_rm_p c^5 T^3}\biggr)^{1/8}\biggl(\frac{1+z}{2}\biggr)^{3/8}\cr 
&=&255.2\,\biggl(\frac{1+z}{2}\biggr)^{3/8}\,n^{-1/8}_{r,1}\,E^{1/8}_{54}\,\biggl(\frac{T_{90}}{32s}\biggr)^{-3/8}
\eary
where $T_{90}$ is the time of the GRB, which is much larger that the peak time of the reverse shock emission, and $n_r$ is the thick shell density. We consider the thick shell case in which the reverse shock becomes relativistic during the propagation and the shell is significantly decelerated by the reverse shock.  Hence, the Lorentz factor at the shock crossing time $t_c$ is given  by $\gamma_d\sim \Gamma_c$ \cite{kob07a,kob07b} and for $\sigma=L_{pf}/L_{kn}\sim 1$ the crossing time $t_c$ is much shorter  than $T_{90}$, $t_c\sim T_{90}/6$,  \cite{fan04a, fan08, zha05, dre02}.   Now, if the constant fractions, $\epsilon_{e,r}$ and $\epsilon_{B,r}$ of the reverse shock energy go into the electrons and magnetic fields, respectively, we have
\bary
\gamma_{\rm m,r}&=&\epsilon_{e,r}\biggl(\frac {p-2} {p-1}\biggr) \frac {m_p} {m_e}\frac{\gamma_r}{\Gamma_c}\cr
&=& 1233.5\,\biggl(\frac{1+z}{2}\biggr)^{-3/8}\,\biggl(\frac{\epsilon_{e,r}}{0.6}\biggr)\,\gamma_{\rm r,3}\,n^{1/8}_{r,1}\,E^{-1/8}_{54}\,\biggl(\frac{T_{90}}{32s}\biggr)^{3/8}
\eary
where $\gamma_r$ is the Lorentz factor of the thick shell. The spectral characteristics of the forward and reverse shock synchrotron emission are related \cite{zha03,kob07a,fan05,fan04a,jin07,sha05} by,

\bary\label{conec}
\nu_{\rm m,r}&\sim&\,\mathcal{R}^2_e\,\mathcal{R}^{-1/2}_B\,\mathcal{R}^{-2}_M\,\nu_{m,f}\cr
\nu_{\rm c,r}&\sim&\,\mathcal{R}^{3/2}_B\,\mathcal{R}^{-2}_x\,\nu_{c,f}\cr
F_{\rm max,r}&\sim&\,\mathcal{R}^{-1/2}_B\,\mathcal{R}_M\,F_{max,f}
\eary
where $\mathcal{R}_B=\epsilon_{B,f}/\epsilon_{B,r}\,$, $\mathcal{R}_e=\epsilon_{e,r}/\epsilon_{e,f}\,$,  $\mathcal{R}_x=(1+x_f)/(1+x_r+x_r^2)$ and $\mathcal{R}_M=\Gamma^2_c/\gamma$.  The previous relations tell us that including the re-scaling there is a unified description between both shocks (forward and reverse).  Such as the magnetic field where there are some central engine models\cite{uso92, mes97b, whe00} for which the fireball wind may be endowed with "primordial" magnetic fields. Also as the cooling Lorentz factor must be corrected, then  $\mathcal{R}_x$ is introduced as a correction factor for the IC cooling, where $x_r$ is obtained by \cite{kob07a} as,

\begin{equation}
x_r = \left\{ \begin{array} {ll} 
\frac{\eta \epsilon_{e,r}}{\epsilon_{B,r}}, & \mathrm{if \quad}
\frac{\eta \epsilon_{e,r}}{\epsilon_{B,r}} \ll 1, \\ 
\left(\frac{\eta \epsilon_{e,r}}{\epsilon_{B,r}}\right)^{1/3}, & \mathrm{if \quad}
\frac{\eta \epsilon_{e,r}}{\epsilon_{B,r}} \gg 1. 
\end{array} \right.
\end{equation}
Using equations (\ref{synforw}) and (\ref{conec}), the typical and cooling frequencies of the reverse shock synchrotron emission are
\bary\label{synrev}
\nu_{\rm m,r}&\sim& 3.4\times 10^{16}\biggl(\frac{1+z}{2}\biggr)^{-1}\,\biggl(\frac{\epsilon_{e,r}}{0.6}\biggr)^{2}\,\biggl(\frac{\epsilon_{B,r}}{0.125}\biggr)^{1/2}\cr
&&\gamma^{2}_{r,3}\,n^{1/2}_{r,1}\, {\rm \ Hz}, \cr
\nu_{\rm c,r}&\sim& 1.5\times 10^{11} \biggl(\frac{1+z}{2}\biggr)^{3/2}\,\biggl(\frac{1+x_r+x_r^2}{6}\biggr)^{-2}\cr
&&\biggl(\frac{\epsilon_{B,r}}{0.125}\biggr)^{-7/2}n^{-3}_{r,1}\,E^{-1/2}_{54}\,\gamma^{-6}_{r,3}\,\biggl(\frac{T_{90}}{32s}\biggr)^{5/2}\, {\rm \ Hz}, \cr
F_{\rm max,r}&\sim& 5.02\times 10^2 \biggl(\frac{1+z}{2}\biggr)^{7/4}\,\biggl(\frac{\epsilon_{B,r}}{0.125}\biggr)^{1/2} \,n^{1/4}_{r,1}\,D^{-2}_{28}\,E^{5/4}_{54}\cr
&&\gamma^{-1}_{r,3}\biggl(\frac{T_{90}}{32s}\biggr)^{-3/4}\,{\rm \,Jy}.
\eary

From equation~(\ref{synrev}) we see that $\nu_{\rm m,r}$ and $\nu_{\rm c,r}$  correspond to optical and IR frequencies, respectively. However  these energies  were not recorded. Instead, as higher energy photons were observed we compute the upscattering emission of the synchrotron radiations  by relativistic electrons  \cite{sar01}), 
\bary\label{ssc}
\nu^{(IC)}_{\rm m}&\sim& 1.034\times 10^{23} \biggl(\frac{1+z}{2}\biggr)^{-7/4}\,\biggl(\frac{\epsilon_{e,r}}{0.6}\biggr)^{4}\,\biggl(\frac{\epsilon_{B,r}}{0.125}\biggr)^{1/2}\cr
&&\gamma^{4}_{r,3}\,n^{3/4}_{r,1}\,E^{-1/4}_{54}\,\biggl(\frac{T_{90}}{32s}\biggr)^{3/4}\, {\rm \ Hz},\cr
\nu^{(IC)}_{\rm c}&\sim& 1.1\times 10^{10} \biggl(\frac{1+z}{2}\biggr)^{3/2}\,\biggl(\frac{1+x+x^2}{6}\biggr)^{-4}\cr
&&\biggl(\frac{\epsilon_{B,r}}{0.125}\biggr)^{-7/2}\,n^{-3}_{r,1}\,E^{-1/2}_{54}\,\gamma^{-6}_{r,3}\,\biggl(\frac{T_{90}}{32s}\biggr)^{-5/2}\, {\rm \ Hz},\cr
F^{(IC)}_{\rm max}&\sim& 4.7 \times 10^{-2} \biggl(\frac{1+z}{2}\biggr)^{9/4}\,\biggl(\frac{\epsilon_{B,r}}{0.125}\biggr)^{1/2}\,n^{3/4}_{r,1}\,D^{-2}_{28}\cr
&&E^{7/4}_{54}\,\gamma^{-2}_{r,3}\,\biggl(\frac{T_{90}}{32s}\biggr)^{-5/4}\,{\rm \,Jy}.
\eary

From equation~(\ref{ssc}) we observe  the break energies and $(\nu F)_{\rm max}=21.2\times 10^{-6}\,\frac{erg}{cm^2\,s}$ are within the range pointed out by \cite{gon09,gon11}.

\section{Discusion and Conclusions}
Choosing $t_0\leq14$s \citep{gon09,gon11,  sac11a}, and assuming that from this time until $t\sim32$s the synchrotron emission was eclipsed by the prompt, the transition time between fast and slow cooling is $\sim 8.7\,$s, so 18s later  the synchrotron process generated by the forward shock was in the slow-cooling regime and the energy range obtained corresponds to that reported by \cite{Giblin et  al.}. Also, we suggest that  the diminishing flux at $\simeq14$~s may be due to pair production  ($\gamma\gamma\to e^+e^-$) between prompt emission  and  forward shock photons (equation \ref{synforw})  at the beginning of the first second. \\ 
In the reverse shock, the synchrotron process emitted photons with $\nu_{\rm c,r}\sim 1.0\times 10^{10}\,$Hz and $\nu_{\rm m,r}\sim4.6\times 10^{16}\,$Hz, which were not recorded but were  upscattered by electrons  up to break energies $E^{IC}_{\rm c}\sim 4.2\times 10^{-5} \,$eV, $E^{IC}_{\rm m}\sim\,427.9\,$MeV with a  $(\nu F)_{\rm max}=21.2\times 10^{-6}\,\frac{erg}{cm^2\,s}$, which were pointed out by \cite{gon09,gon11}  and  \cite{sac11a}.   Now,  in accordance with the observed value for the power index $\beta\sim 1.44\pm 0.07$,  $\nu^{(ic)}_{m,r}>\nu^{(ic)}_{c,r}$   we conclude the  SSC spectrum corresponds to fast-cooling regime, very similar to GRB 941017 \citep{gon03,gra03}. For our case (thick case), the flare occurs during the prompt gamma-ray phase.\\ 
In accordance with  $\mathcal{R}_B$, we obtained that forward and reverse magnetic fields are related by  $B_f=0.9\times 10^{-3}B_r$.  The previous result indicates  that there was  a stronger magnetic field in the reverse-shock region than in the forward shock region, which may suggest that the obtained  results are given when the ejecta is magnetized, as  in the interpretation of the early afterglow of GRB 990123 and GRB 021211 provided  by \cite{zha03}.\\
Finally, because the Large Area Telescope (LAT) covers the energy range from about 20 MeV to more than 300 GeV,  we hope to detect other hard components in GRBs and so further constrain this model.

The current model accounts for the main characteristics of the burst: energies, spectral indices, fluxes, duration of the main components in a unified manner. The main requirements are that the ejecta be magnetized, leading to the formation of a reverse shock. The model has eight free parameters (equipartition magnetic field, equipartition electron energy, Lorentz factor, and densities all of the in the reverse and forward shocks), with standard values. 

This burst has similar characteristics to GRB 090926A~\cite{ack11}, but the high energy extended emission component may require SSC from the forward shock \citep{sac11b}.

\bigskip 
\begin{acknowledgments}

We thank Enrico Ram\'{\i}rez-Ruiz and  Charles Dermer for useful discussions. This work is partially supported by UNAM-DGAPA PAPIIT IN105211 (MG) and CONACyT 103520 (NF) and 83254 (WL). 

\end{acknowledgments}

\bigskip 

\end{document}